# Spatially Heterogeneous Dynamics in a Metallic Glass Forming Liquid Imaged by Electron Correlation Microscopy

Pei Zhang[1], Jason J. Maldonis[1], Ze Liu[2], Jan Schroers[2], Paul M. Voyles[1*]

1. Department of Materials Science and Engineering, University of Wisconsin-Madison, Madison, WI 53706, USA

2. Department of Mechanical Engineering and Materials Science, Yale University, New Haven, Connecticut 06511, USA.

*Corresponding author.  Email: paul.voyles@wisc.edu

## Abstract

Supercooled liquids exhibit spatial heterogeneity in the dynamics of their fluctuating atomic arrangements. The length and time scales of the heterogeneous dynamics are central to the glass transition and influence nucleation and growth of crystals from the liquid. We report direct experimental visualization of the spatially heterogeneous dynamics as a function of temperature in the supercooled liquid state of a Pt-based metallic glass, using electron correlation microscopy with sub-nanometer resolution. An experimental four point space-time intensity correlation function demonstrates a growing dynamic correlation length, $\xi$, upon cooling of the liquid toward the glass transition temperature. $\xi$ as a function of the relaxation time $\tau$ data are in the good agreement with the Adam-Gibbs, inhomogeneous mode coupling theory and random first order transition theory of the glass transition. The same experiments demonstrate the existence of a nanometer thickness near surface layer with order of magnitude shorter relaxation time than inside the bulk.

The dynamics of the rearrangements of atoms or molecules in a liquid are challenging to determine directly, yet they govern critical processes including diffusion[1], viscous flow[2], nucleation and growth of crystal phases[3], and the glass transition[4,5]. In the supercooled state, it is envisioned that dynamics become spatially heterogeneous[6], with nanoscale domains exhibiting widely varying characteristic structural relaxation times[7–12]. According to most microscopic theories, the relaxation time and characteristic length scale of slow domains grows as the liquid cools through the glass transition[13–16]. In contrast, classical nucleation theories typically treat the liquid prior to the transition to a crystal as a uniform interface[17]. Experimental data on liquid dynamics comes from their frequency dependent response to mechanical, electrical, or thermal stimuli[18–20], spectroscopy[21–23], or scattering of coherent light[24–27]. These are bulk measurements, from which spatial heterogeneity in dynamics largely inferred, with a few indirect exceptions[28]. Here, we apply electron correlation microscopy (ECM)[29,30] which reveals the spatially heterogeneous nanodomains in the supercooled state of a bulk metallic glass forming liquid. The ECM data provide direct evidence of a growing dynamical length scale as the liquid cools toward the glass transition, consistent with the random first order theory of the glass transition[14], and identify a near-surface layer with fast dynamics that influences crystallization[31].

ECM uses time-resolved coherent electron scattering to study liquid dynamics at the nanoscale[29,30]. When performed in diffraction mode with a focused probe, it is the electron equivalent of (x-ray) photon correlation spectroscopy[23,32]. coherent scattering produces a speckle pattern, $I(\mathbf{k}, t)$, where $I$ is the intensity, $\mathbf{k}$ is the scattering vector, and $t$ is time. Each speckle corresponds to a volume of the sample at the length scale of probe size with sufficient internal order to create constructive interference of the scattered waves. Local structural rearrangements cause intensity fluctuations of the speckle, so the lifetime of a speckle represents the time over which a particular structure persists. Statistical analysis of many speckles using the time autocorrelation function

$$g_2(t) = \frac{\langle I(t')I(t'+t)\rangle}{\langle I(t')\rangle^2}, \tag{1}$$

where *t'* is the time of a frame in the diffraction time series, *t* is delay time after *t'*, and <> denotes average over all *t'*, fit to the Kohlrausch-Williams-Watt (KWW) equation

$$g_2(t) = 1 + A exp[-2\left(\frac{t}{\tau}\right)^\beta],  \quad (2)$$

yields the characteristic relaxation time $\tau$ and the stretching exponent $\beta$[33]. The advantage of ECM is spatial resolution. A modern transmission electron microscope (TEM) can create sub-nanometer diameter beams with high coherence[29,34], and the electron's large scattering cross section creates nanodiffraction patterns with acceptable signal to noise ratio even from small volumes.

Previous ECM experiments were performed using a stationary, nanometer diameter probe beam, so they measured $I(\mathbf{k}, \mathbf{r}, t)$ at only one point **r** on the sample. Figure 1(a) shows how ECM implemented using dark-field TEM imaging can measure many positions in parallel, producing an image of the entire sample of the liquid state dynamics. A broad coherent electron beam illuminates the sample, and a diffraction pattern $I(\mathbf{k}, t)$ is formed in the back focal plane of the objective lens. A small objective aperture is introduced, blocking most of the pattern, but admitting a speckle or two at a particular **k**. Additional optics form a real-space image from just the scattered electrons in those speckles, producing a spatial map of the speckle intensity, $I(\mathbf{k}, \mathbf{r}, t)$. This represents the same basic data as probe-based ECM, but with one point in **k** instead of many, and many points in **r** instead of one. (The difference is exactly analogous to the difference between probe-based and dark-field image based fluctuation electron microscopy[35].) A time series of images enables us to track the speckle intensity at a particular position on the sample, revealing the dynamics at that position.

ECM experiments were performed on $Pt_{57.5}Cu_{14.7}Ni_{5.3}P_{22.5}$ nanowires[36] in a FEI Titan microscope operated at 200 kV with a 10 μm objective aperture with a diameter in reciprocal space of 2.51 nm$^{-1}$, corresponding to a real-space image resolution of 0.7 nm. The nanowires are small enough to be used as TEM samples without requiring thinning with ion beams that can damage the sample surface, creating catalytic sites for crystallization. The wires were heated at the rate of 20 K/min into the supercooled liquid state between the glass transition temperature $T_g$ = 507 K up to 523 K, near the crystallization temperature[37], using a DENS Solutions MEMS-based heating holder with 0.1 K temperature stability and <5 K absolute temperature accuracy. Before recording the image series, the nanowires were equilibrated at the elevated temperature for at least five times the median relaxation time $\tau_{med}$ (See supplementary text for the additional details). At higher temperatures, the nanowires crystallized before a long enough time series could be acquired. Image time series were acquired on a Gatan Orius CCD camera. The interval between frames was set as a function of temperature to ~$0.005\tau_{med}$, where $\tau_{med}$ was acquired with a time interval short enough not to influence the results[30]. The total time for the time series was set to ~$20\tau_{med}$, to provide a balance between a time series that is too long, which may average together short and long relaxation process over temporally fluctuating dynamics, yielding artificial spatial homogeneity in the relaxation time[38], and a time series that is too short, which yields autocorrelation functions that are not well converged[10,39]. $20\tau_{med}$ can be thousands of seconds near $T_g$, so rigid image alignment was used to correct sample drift (see supplementary video 1 and related supplemental information). Figure 1(b) shows an example of the normalized, resampled $g_2(t)$ calculated from the intensity fluctuations of one pixel inside a nanowire at 523 K. It shows good convergence to zero at long times and follows the KWW stretched exponential behavior. Additional information about the sample preparation, data acquisition, and data analysis may be found in the supplemental text.

Figure 2 shows representative images of $\tau$ as a function of temperature, calculated from $g_2(t)$ at every pixel. The images show striking visual evidence of nanoscale spatially heterogeneous dynamics in the supercooled liquid. At higher temperature, like Figure 2(a) ($T_g$ + 16 K / 523 K), $\tau$ varies by an order of magnitude from place to place, with an apparently random distribution of domains ~1 nm in diameter. As the temperature approaches $T_g$, the $\tau$ distribution becomes wider, covering two orders of magnitude near $T_g$ (Figure 2(e), 507 K). The domains with similar $\tau$ also appear to grow in size, with more globular, extended domains several nanometers in diameter developing. At all temperatures, the histogram of relaxation times is well described by a log-normal distribution, characteristic of a random, non-negative process (see supplemental Figure S9), the width of which increases with decreasing temperature. The stretching exponents β range from 0.2 to 1. β < 1 indicates a superposition of multiple relaxation processes with different characteristic times, which is typical for complex liquids near $T_g$ (e.g. [23] for a metallic glass). With increasing temperature, rearrangements in liquids become less correlated and eventually only one relaxation process prevails, mathematically reflected by β = 1. The observed decrease of the mean β with decreasing temperature approaching $T_g$ (see supplemental Figure S9), indicates more dynamic heterogeneity, in line with the direct observation (Figure 2). The temporal evolution of the domains is visualized using a sliding window along the time series in supplemental video 2.

In the column approximation, each pixel in Figure 2 arises from a cylinder of material with a diameter of 0.7 nm, extending through the thickness of the sample. However, not all of that material contributes to the intensity in the image at every time. Only a small portion of each cylinder contains atoms arranged in such a way that they scatter into the ~1 speckle that passes through the objective aperture and contributes to the image. Most of the atoms either do not have constructive interference of their scattering at all, or they scatter in a different direction into a different speckle that is blocked by the aperture. We estimate that the number of ordered clusters scattering through the aperture from one column is between 1 and 2 for the thickness of nanowires used here (see supplemental information). Experimentally, the profiles of τ as a function of position in from the edge of the nanowire (and thus increasing thickness) in Figure 3 and Figure S5 and S6, show that there is a systematic variation in τ for small thicknesses, but we can identify a region in the center of each nanowire where the thickness effects are negligible. Figure S4 shows that in this central region, the measured τ is independent of the sample thickness for thicknesses between 28 and 45 nm, including the 40 nm thick wires analyzed more extensively. Finally, Figure S6 is a τ image of a rectangular cross-section nanowire with uniform thickness, demonstrating that the observed fluctuations in τ are not caused by local thickness variations.

The average τ(T) derived from central portion of the images, shown in Figure 2(f), is in reasonable agreement with bulk measurements on the same alloy. τ extracted from the bulk increases from ~20 s to ~350 s when the temperature is cooled from $T_g$ + 16 K to $T_g$, which is consistent with data reported for supercooled metallic glass forming liquids[20,40–42]. $\tau(T)$ follows the Arrhenius law over the limited temperature range accessible in ECM, with an activation energy of 3.7 ± 0.3 eV, in good agreement with typical values for bulk metallic glass alloys[43,44], and the viscosity estimated from the Debye-Stokes Einstein equation is in reasonable agreement with estimates in the literature for the same alloy[45].

Quantitative characterization of spatially heterogeneous dynamics in simulations has been accomplished using high order correlation functions[46–50]. We have calculated a similar two-time, two-position correlation function to derive a characteristic dynamic correlation length $\xi$

$$g_4(\Delta t, \Delta r) = \frac{\sum_{t,r} I(t,r)I(t+\Delta t,r)I(t,r+\Delta r)I(t+\Delta t,r+\Delta r)}{\sum_{t,r} I(t,r)\sum_{t,r} I(t+\Delta t,r)\sum_{t,r} I(t,r+\Delta r)\sum_{t,r} I(t+\Delta t,r+\Delta r)}, \quad (3)$$

in which $I$ is the intensity of position $\vec{r}$ in the image, $t$ is the time of a frame in the diffraction time series, and $I(t, r+\Delta r)$ is the average intensity from all the pixels a distance $\Delta r$ from position $\vec{r}$. In simulation, functions like $g_4$ are calculated for the same particle(s) as a function of time, which we cannot do in experiments. However, ensemble averaging over many positions and long times converges to the same shape of function, as is well known for $g_2(t)$, the self-intermediate scattering function (in simulations)[51] and the intermediate scattering function (in experiments)[52].

Figure 4(a) shows an example $g_4(\Delta t, \Delta r)$ at 507 K. We extracted the spatial decay of $g_4$ by fixing the $\Delta t = \tau(T)$ and obtained the correlation length $\xi$ through the exponential fitting as described in supplemental Figure S1 and accompanying text. We also extracted $\xi$ and a characteristic time $\nu$ simultaneously from the two-dimensional shape of $g_4$ as described in supplemental Figure S2. The results are very similar and the conclusions are the same (see Supplemental Figure S3 and accompanying discussion).

$\xi(T)$ increases from ~0.82 nm at 523 K to ~1.41 nm at 507 K, as shown in Figure 4(b). This range of length scales and general behavior is broadly consistent with previous theoretical and experimental estimates[11,15,53]. However, $\xi$ at the temperature closest to $T_g$ does not increase as strongly as expected. Further experiments with better temperature stability, longer data series, and potentially other alloys are required to determine if this single point represents new physics or if it is some limitation of the current experiments. Various microscopic models of dynamics in the supercooled liquid and the glass transition predict different behavior for $\xi(\tau)$. Figure 4(c) shows the linear fit between $\ln(\tau)$ with $\xi$, predicted by the Adam-Gibbs theory of cooperatively rearranging regions[15]. The data follow the prediction well, except for the data nearest to $T_g$ (longest $\xi$ and $\tau$). The inhomogeneous mode coupling theory (IMCT)[54] predicts a power-law relationship. Figure 4(d) shows a log-log plot of $\xi(\tau)$, which is also a good fit at high $T$ but deviates close to $T_g$. The random first order transition (RFOT) theory[14] predicts $\tau \sim \exp(k(\xi/k_BT)^n)$. Figure 4(e) shows $\ln(\tau)$ as a function of $\xi/k_BT$. A line fits over the entire data range. The reduced $\chi^2$ for all three fits is similar (0.914 for Adam-Gibbs, 0.905 for IMCT and 0.961 for ROFT due to the additional fitting parameter), so we cannot draw firm conclusions on the correct model at this time. Future experiments over a wider range of $\tau$ and $T$ may provide further insight.

The other striking feature in Figure 2 is a near-surface layer with much faster dynamics than present in the bulk. The width of the near-surface layer is ~1 nm, as shown in Figure 3, and $\tau$ varies from a few seconds at high temperature to ~20 s near $T_g$. Surface diffusion in solids is often faster than bulk diffusion, both for crystals[55] and for glasses[31,56,57], and similar observations have been made in liquids[58]. However, the near-surface layer observed here involves ~1014 atoms, of which only ~200 are in the first atomic layer in contact with vacuum, so it is unlikely that surface diffusion dominates. Nor is the layer an effect of the shape of the wire, since a similar fast near-surface layer is observed on a rectangular wire (Supplemental Figure S6). From Figure 2(f), the activation energy for the near surface layer is $1.7 \pm 0.3$ eV, consistent with the prediction from the RFOT theory that the free energy barrier for activated motion near a free surface should be half that of the bulk[59]. A similar highly mobile but significantly thicker surface layer has been proposed as an explanation for the suppression of $T_g$ in very thin polymer films.[60]

It has been widely observed that crystallization typically proceeds inwards from the surface[61], a phenomenon that is usually attributed to heterogeneous nucleation on the sample's surface and quantified by a reduction in the activation volume $f(\theta) = [0,1]$[62,63]. Surface initiated crystallization has been observed on nanowires similar to those studied here, and it has been used to explain the strong dependence of the crystallization rate on the diameter of the nanowire[64]. The enhanced surface mobility shown in Figure 2 and Figure 3 provides an alternative explanation for surface crystallization. To estimate the effect that the enhanced surface mobility has on the nucleation rate, we compare the classical homogenous nucleation rate in the surface layer with enhanced mobility and to the heterogeneous nucleation rate with a catalytic surface site. The classical homogenous nucleation rate can be estimated by $I_{surfacemobility} \propto A/\eta_{surface} N_{surfacelayer} \exp(-\Delta G^*/kT)$, in which $I_{surfacemobility}$ is the nucleation rate in the ~1 nm thick near-surface layer, A is a constant, $\eta_{surface}$ is the viscosity in the surface layer, $N_{surfacelayer}$ is the number of atoms in the surface layer, and $\Delta G^*$ is the activation energy for the formation of a stable nuclei. The heterogeneous nucleation rate is $I_{Het} \propto A/\eta N_{surface} \exp(-\Delta G^* f(\theta)/kT)$, in which $N_{surface}$ is the number of surface atoms and $\eta$ is the bulk viscosity. The condition $I_{surfacemobility} = I_{Het}$ allows to calculate the required catalytic heterogeneous influence, quantified by $f(\theta)$, to match homogeneous nucleation with a low-viscosity surface layer. With the parameters and temperature considered here, the rates are equal at $f(\theta) = 0.5$. When we consider non-classical models, which have been shown to be more appropriate in describing nucleation of deeply undercooled melts[65,66], such as present here, $I_{surfacemobility}$ approaches $I_{Het}$ with $f(\theta) \sim 1$. This suggests that enhanced surface mobility can explain widely observed surface nucleation equally well as heterogeneous nucleation. As the main evidence for heterogonous nucleation is the observation that nucleation proceeds from the surface, the results presented here suggest that this may need to be reconsidered, and that more sophisticated experiments are required to reveal the origin of widely observed surface nucleation in metals.

In summary, we demonstrated direct measurement at the nanoscale of spatial heterogeneous dynamics in the supercooled liquid state of Pt-based alloy nanowires using dark-field electron correlation microscopy. The dynamics are characterized by a growing length and time scale as the liquids cools toward the glass transition. The nanowires also exhibit a near-surface layer ~1 nm thick with substantially faster dynamics than the bulk. The near surface layer provides an effective mechanism for surface crystallization of liquids by homogeneous, as opposed to heterogeneous, nucleation.

# Acknowledgments

We thank Mark Ediger for helpful discussions. This work was supported by the US National Science Foundation under Contract No. DMR-1506564 (P. Zhang and P. M. Voyles). The facilities and instrumentation for TEM sample preparation and microscopy at UW-Madison were supported by the University of Wisconsin Materials Research Science and Engineering Center (DMR-1121288). Ze Liu and Jan Schroers were supported by the Department of Energy through the Office of Basic Energy Sciences (#DE SC0004889).

Figures and Captions

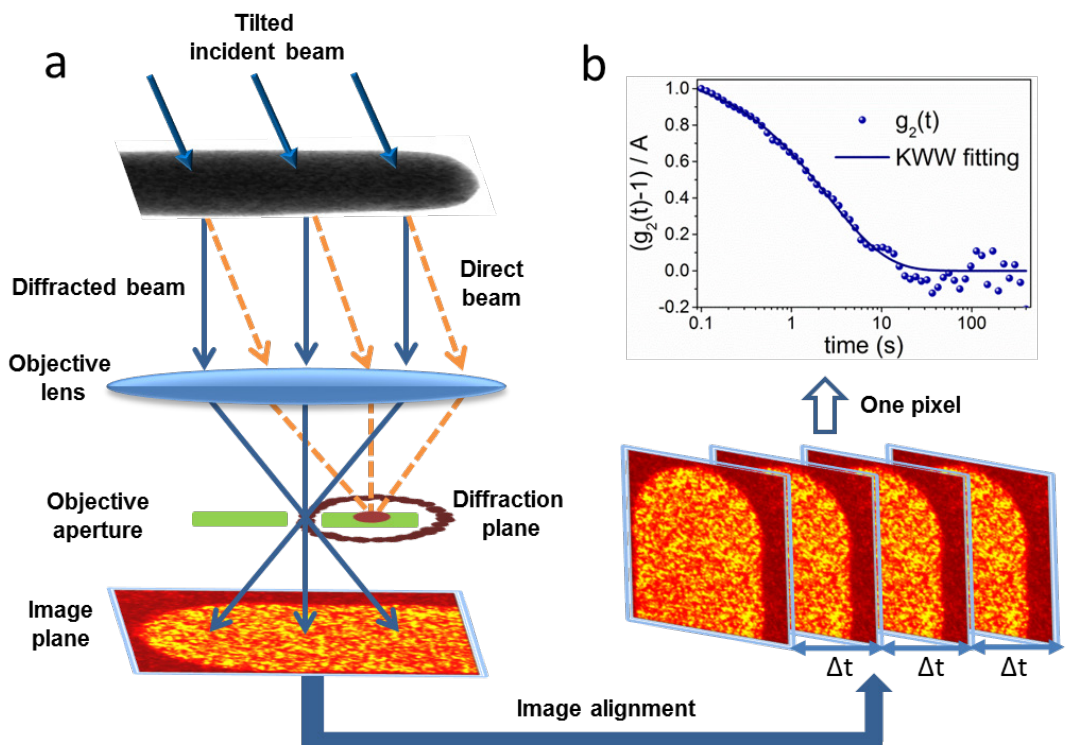

Figure 1. **Tilted dark-field electron correlation microscopy** (a) Schematic of the experiment. Tilted illumination shifts the transmitted beam off the optic axis of a microscope. A small, on-axis objective aperture selects one speckle in the diffraction pattern, forming a real-space image. Structural rearrangements cause fluctuations in the intensity of the image speckles. A time series of ~4000 dark field images is recorded and then aligned to correct drift of the sample. (b) The time autocorrelation function $g_2(t)$ is calculated from the intensity time series at every pixel, then fit to the KWW equation to extract the relaxation time τ and stretching parameter β. The figure shows a typical $g_2(t)$ which is well converged. This calculation is repeated for every pixel in the image.

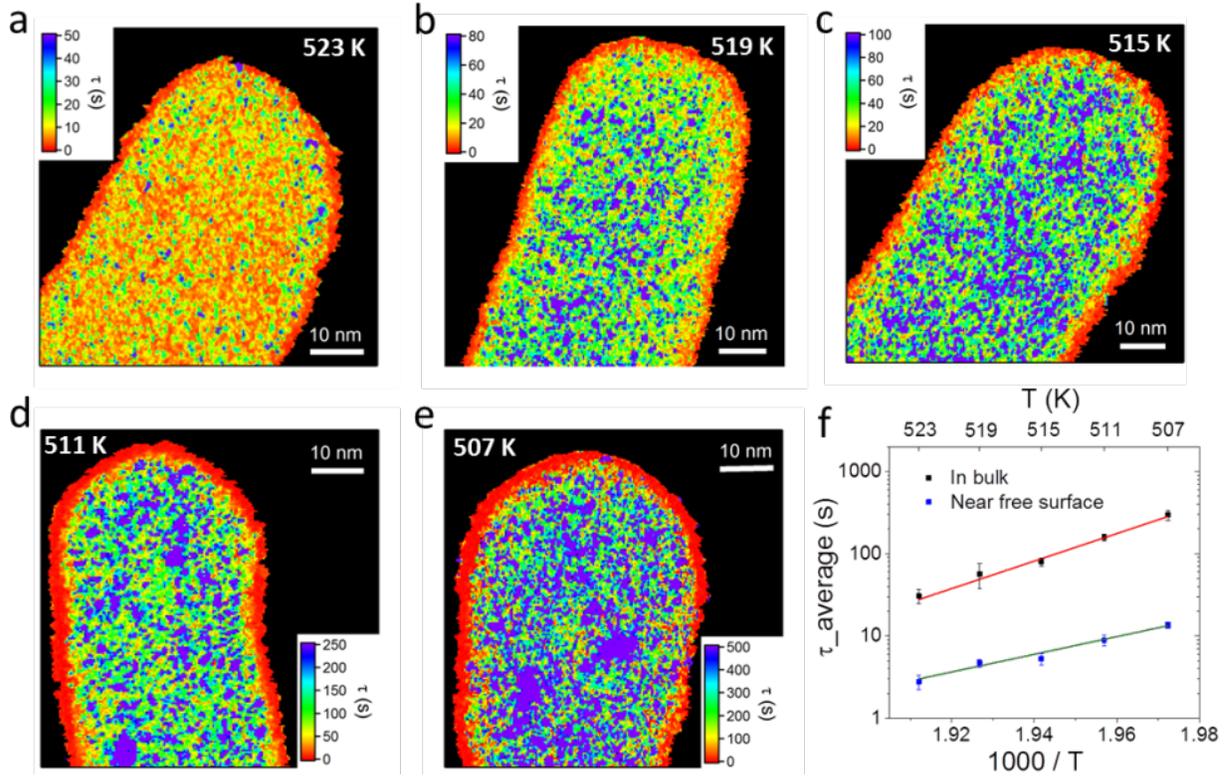

Figure 2. **Spatial maps of structural relaxation time on the nanowire in the supercooled liquid regime.** (a) $T$ = 523 K, (b) $T$ = 519 K, (c) $T$ = 515 K, (d) $T$ = 511 K, (e) $T$ = 507 K. The maps show domains with varying relaxation time at the nanometer scale. With decreasing temperature, slow domains appear larger and occupy a greater fraction of the map, especially very close to $T_g$ = 507 K. There is a region ~1 nm thick with ~20 times shorter relaxation time near the surface of every wire. (f) The mean structural relaxation time for the nanowire interior (bulk) and the near-surface layer. The error bars are the standard deviation of the mean of four measurements on different nanowires. Fitting to the Arrhenius form yields activation energies of 3.7 ± 0.3 eV for the bulk and 1.7 ± 0.3 eV for the near-surface.

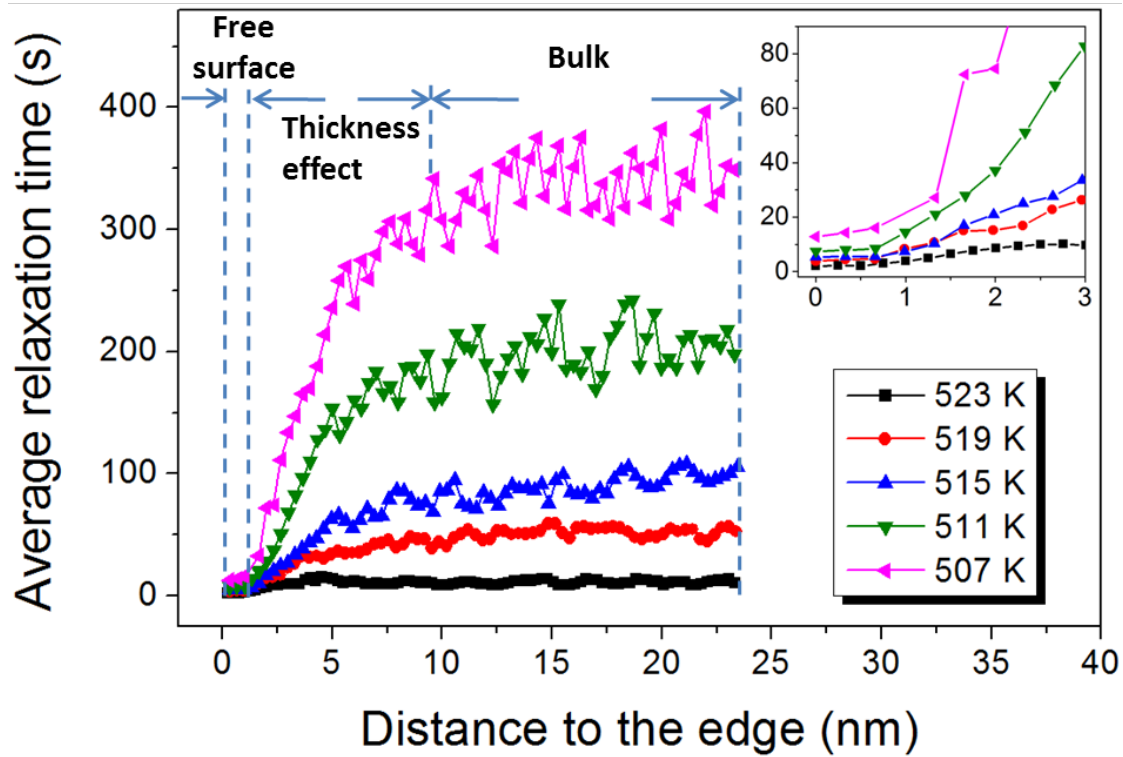

Figure 3. **Near surface region with fast dynamics**. Profiles through the t maps in Figure 2. Near the free surface, the relaxation time remains almost constant and is about 20 times smaller than in the center of the nanowire. The inset shows that the near free surface layer is 3-4 pixels wide, which is between 0.7 nm and 1 nm. The dependence of t on the projected thickness of the wire along the profile is discussed in the supplemental information.

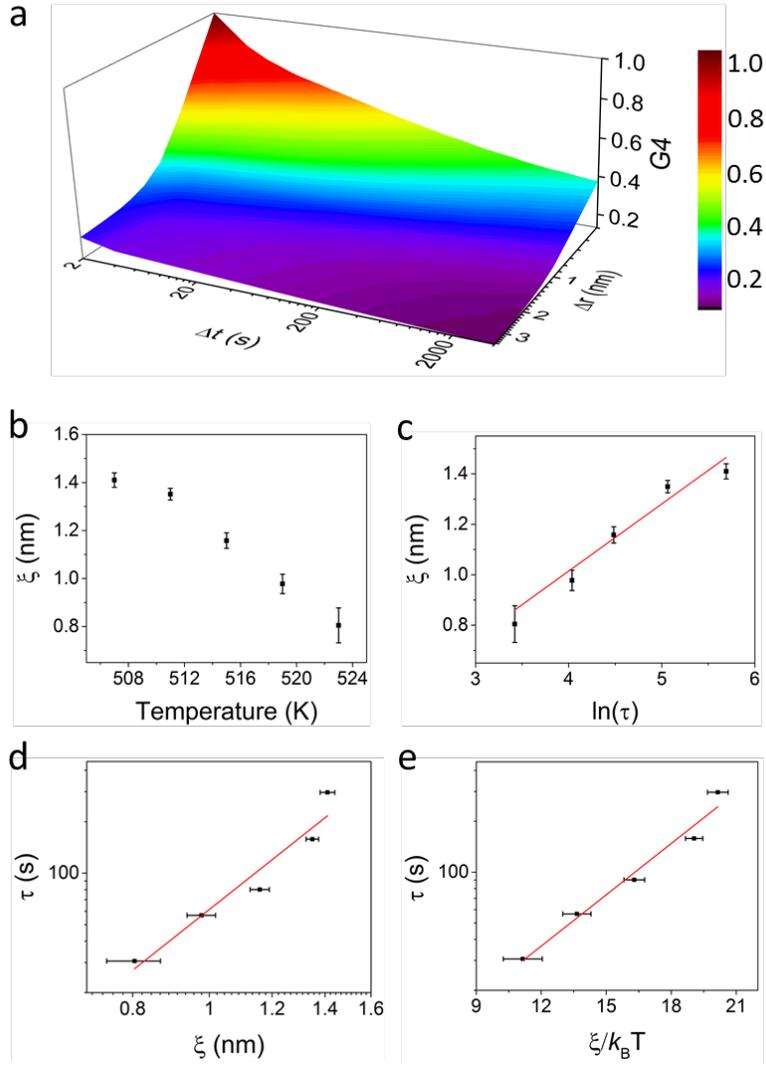

Figure 4. **Characteristic length and time scale for heterogeneous dynamics.** (a) An example $g_4(\Delta t, \Delta r)$ at 507 K. (b) The correlation length ξ as a function of temperature. (c) ξ vs. ln(τ), which is a straight line in the Adam-Gibbs theory.[15] (d) A log-log plot of ξ vs τ, which is a straight line in the inhomogeneous mode coupling theory.[54] (e) A semi-log plot of τ vs ξ/$k_B$T, which is a straight line in the random first-order transition theory.[14] The data are in reasonable agreement with all of the models. The reduced $\chi^2$ is 0.914, 0.905 and 0.961 for the fitting in (c)(d)(e) respectively.

Supplementary Materials for
Spatially Heterogeneous Dynamics in a Glass Forming Liquid
Imaged by Electron Correlation Microscopy

Pei Zhang[1], Jason J. Maldonis[1], Ze Liu[2], Jan Schroers[2], Paul M. Voyles[1]*

1. Department of Materials Science and Engineering, University of Wisconsin-Madison, Madison, WI 53706, USA

2. Department of Mechanical Engineering and Materials Science, Yale University, New Haven, Connecticut 06511, USA.

*Corresponding author. Email: paul.voyles@wisc.edu

**Materials and methods**

1. Sample preparation

$Pt_{57.5}Cu_{14.7}Ni_{5.3}P_{22.5}$ glassy nanowires with diameter of 40~45 nm were fabricated by the nanomoulding method described in detail elsewhere.[1] As-fabricated, the nanowires were attached to a substrate plate of the same bulk metallic glass. The metallic glass plate was rinsed with distilled water and isopropyl alcohol to minimize the residual salts and anodized aluminum oxide from fabrication. Then the plate was immersed in methanol and nanowires were released by sonication for 15~20 mins. The methanol containing nanowires was dropped onto the DENS chip through a micro pipet (1.5 ~1.8 μL). After evaporation of methanol, some nanowires were randomly attached on the $SiN_x$ membrane of chip window, which is ~90% electron transparent. This process was repeated several times based on the density of nanowires in methanol to ensure that enough nanowires were attached and isolated. Some contamination was introduced, during sample preparation, probably from the methanol, so the DENSchip was plasma cleaned at 20 psi Ar + $O_2$ mixture for 12 ~ 15 mins before microscopy experiments.

2. Dark field electron correlation microscopy

Previously, ECM experiments were performed using a nanometer-diameter coherent probe beam of electrons to acquire the structural relaxation time at one position on the sample.[2,3] Here, in order to capture spatial maps of the relaxation dynamics, ECM measurements were carried out using tilted dark field TEM imaging. Experiments used the University of Wisconsin-Madison FEI Titan with probe aberration corrector at 200 kV, operated in TEM mode. An objective aperture of 10 μm in diameter or 2.83 mrad of half angle was inserted, giving rise to speckles in the image ~0.7 nm in diameter, calculated from the Rayleigh criterion and confirmed by imaging. The speckle size sets the spatial resolution of the ECM experiment. 256 by 256 pixel images were acquired with the magnification adjusted to yield a typical pixel size of 0.25 nm, so each speckles covers ~3 pixels.

*In situ* heating was done using a DENSsolutions SH30 single-tilt heating holder, which can provide temperature stability of 0.1 K and temperature accuracy of 2%. The temperature range for this measurement is from glass transition temperature $T_g$ to $T_g$ + 16 K, or 507 K to 523 K. Nanowires start to crystalize from the free surface when temperature increased to 527 K. The sample was equilibrated at temperature before data collection. At temperatures of 515 K, 511 K and 507 K, the sample was heated to target temperature at a rate of 20 K/min, and then held for 30-60 mins before image time series acquisition. At 523 K and 519 K, in order to avoid crystallization, the sample was first heated to 508 K at rate of 20 K/min and held for 30-60 mins, then heated to target temperature at the same heating rate and held isothermally for 2 min before data collection. At all temperatures, the equilibration time before data acquisition was at least 5 times the measured structural relaxation time shown in Fig. 2. An Orius 2.6k x 4k fast CCD with 1 ms readout time was used to record the time series of images. The acquisition time per frame was set to 0.1 s, 0.25 s, 0.5 s, 1 s, and 2 s for temperatures 523 K, 519 K, 515 K, 511 K and 507 K respectively. Every image series consists of ~4000 frames.

3. Image alignment

Video 1(a) is movie of a typical image series with 4000 frames acquired at 523 K. The bright spot in the image arises from a crystallized chunk of nanowire, the intensity of which does not change over the whole image series. In addition, because the background scattering intensity from $SiN_x$ is quite small compared with intensity from sample, the edge of nanowire can be well defined. Therefore, with the bright spot and the outline of nanowire as references, drift correction can be realized through rigid image alignment with single pixel precision using the convolution-based alignment in DigitalMicrograph software. Video 1(b) shows the series in Video 1(a) after alignment. The intensity of the pixels fluctuates, but the outline of the nanowire is stationary. Drift correction enables acquisition and analysis of long data series at low temperatures.

4. Time autocorrelation function

$g_2(t)$ was calculated from the intensity time series at a single pixel $I(i)$ from

$$g_2(p) = \frac{(N-p)\sum_{i=0}^{N-p-1} I(i)I(i+p)}{[\sum_{i=0}^{N-p-1} I(i)][\sum_{i=0}^{N-p-1} I(i+p)]}, \tag{1}$$

where $N$ is the total number of frames in the diffraction time series, and $p$ and $i$ indicate position in the time series from 1 to $N$. $g_2(t)$ is related to intermediate scattering function $f(t)$ in an ergodic system by

$$g_2(t) = 1 + C[f(t)]^2, \tag{2}$$

where $C$ is an instrument-dependent parameter.[4]

Structural relaxation kinetics and the intermediate scattering function in relaxation phenomena which originate from a superposition of numerous and different relaxation processes such as in amorphous materials are generally described with the Kohlrausch-Williams-Watt (KWW) function,

$$f(t) = f(0)\exp\left(-\frac{t}{\tau}\right)^\beta, \tag{3}$$

where $\tau$ is the structural relaxation time, and $\beta$ is the stretching exponent. Combining Equations (2) and (3), the relationship between $g_2(t)$ and $\tau$ is

$$g_2(t) = 1 + A\exp[-2\left(\frac{t}{\tau}\right)^\beta], \tag{4}$$

where $A$ is an instrument-dependent scaling parameter related to the partial coherence of the illumination. To extract maps of τ and β, $g_2(t)$ calculated from Eq. (2) for every pixel, resampled logarithmically in time, then fit to Eq. (4) using standard non-linear least-squares fitting.

5. Correlation length from the four point correlation function

Figure S1(a) shows a typical four point correlation $g_4(\Delta t, \Delta r)$ calculated from the inner bulk of nanowire at 507 K. Figure S1(b) shows $g_4(\Delta t = \tau, \Delta r)$, where τ is the average relaxation time calculated from $g_2(t)$ from the same dataset. For this example, $\Delta t = \tau = 320$ s. We fit $g_4(\Delta t = \tau, \Delta r)$ to

$$g_4(\Delta r) = A exp\left[-\frac{\Delta r}{R}\right] + B, \tag{5}$$

to obtain the correlation length $\xi = 2R$ associated with relaxation time τ. ($R$ is a radius and characteristic length is usually thought of as a diameter.) The linear background is an artifact of the geometry of the data, in which we have data in two dimensions $x$, and $y$ in the image, but not in $z$ along the beam direction. The fitting in Figure S1(b) results in a correlation length $\xi$ ~1.42 ± 0.02 nm. Same fitting method was applied to all the other temperatures to obtain the results in the main text Figure 4.

6. Characteristic length and time extracted simultaneously from $g_4$

An alternate approach to analyzing $g_4(\Delta t, \Delta r)$ is to account simultaneously for the decay in both dimensions and by doing so to extract both a characteristic length and a characteristic time from a single function. This approach is useful for our data because the functional form of $g_4(\Delta t, \Delta r)$ in Eq. 3 in the main text does not reduce to $g_2(t)$ for $\Delta r = 0$, so τ derived from $g_2(t)$ is not necessarily the most appropriate time scale.

Figure S2(a) is a contour plot of the same data in Figure S1(a), and Figure S2(b) shows the 0.5 contour. The contour in Figure S2(b) is fit to the sum of a decaying exponential and a linear background. ξ (twice the characteristic length ε) and ν are determined by the width of exponential part, as shown in Figure S2(c), resulting in this case in ξ ~1.44 nm and ν ~14 s. ν derived in this method is not the same as τ derived from $g_2(t)$, so the similar of ξ derived by the two methods shows that ξ is depends on very weakly on Δ*t*.

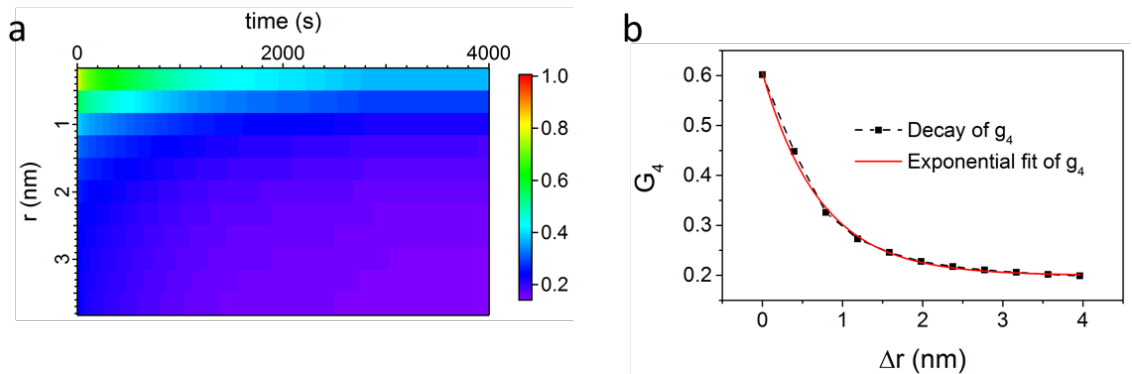

Figure S1. (a) The normalized four-point space time intensity correlation function at 507 K. (b) $g_4(\Delta r)$ with $\Delta t$ fixed at relaxation time τ calculated from $g_2(t)$. The solid line is the exponential fit to the spatial decay of $g_4(\Delta r)$.

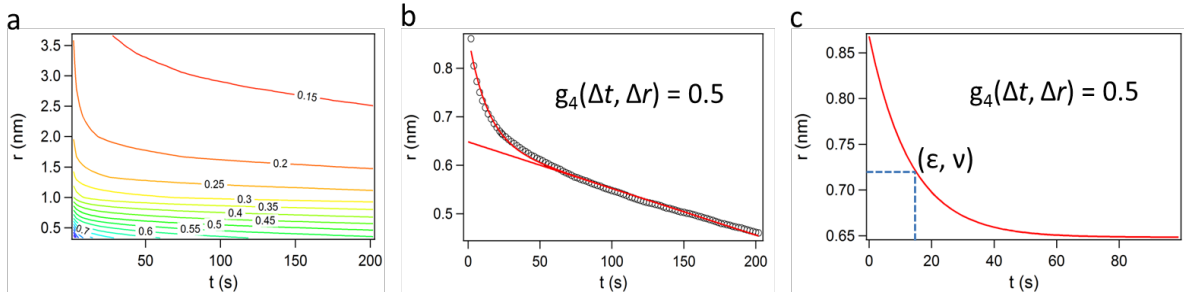

Figure S2. (a) A contour plot of the same data in Figure S1(a). (b) The exponential and linear fit for the (Δ*t*, Δ*r*) contour at $g_4(\Delta t, \Delta r) = 0.5$. (c) The characteristic length ε and a characteristic time ν determined from the width of exponential fitting curve.

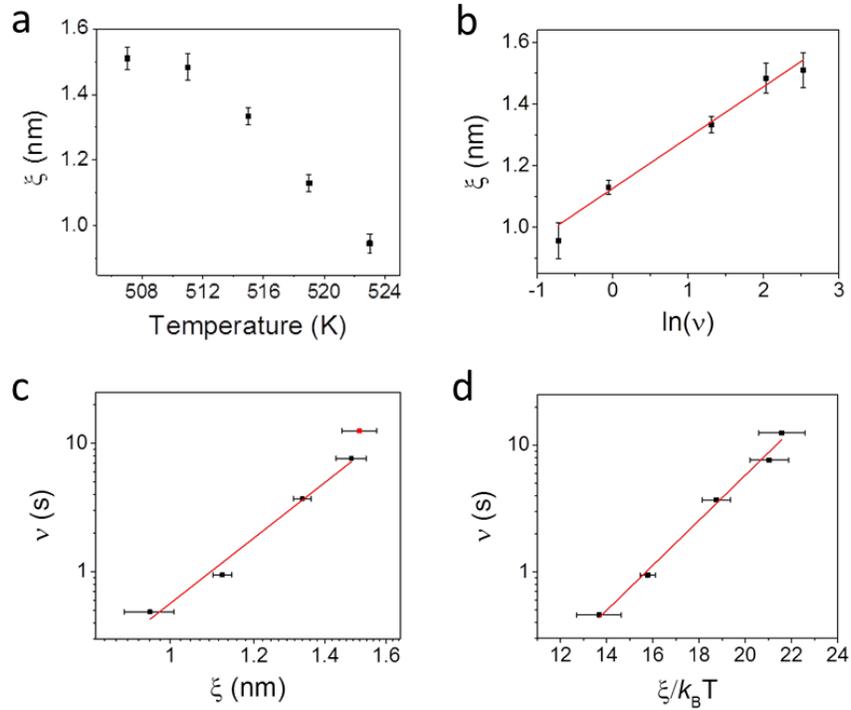

Figure S3. (a) The correlation length ξ as a function of temperature. (b) ξ vs. ln(ν), fit to Adam-Gibbs theory. (c) A log-log plot of ξ vs ν, fit to the inhomogeneous mode coupling theory. (d) A semi-log plot of ν vs ξ/$k_B$T, fit to the random first-order transition theory. The data are in reasonable agreement with all of the models.

Figure S3(a) shows that ξ(*T*) derived in this way increases from ~0.95 nm at 523 K to ~1.5 nm at 507 K, which is around 0.1 nm systematically higher than results derived using the method in Figure S1 and presented in the main text in Figure 4. However, the temperature dependence is very similar. Figure S3(b) (c) (d) shows correlation between ξ with ν, fitted based on the Adam-Gibbs theory, the inhomogeneous mode coupling theory (IMCT) and the random first order transition (RFOT) theory. Reduced $\chi^2$ is similar for the fits to all the models.

**Additional results**

1. Thickness dependence

All TEM experiments measure a projection of the sample along the electron beam direction, so there is the potential for the thickness of the sample to affect the results. For amorphous materials, the strong dynamical diffraction that dominates electron diffraction from crystals does not occur, but inelastic and plural elastic scattering both increase with increasing thickness. The diameter of the nanowires used for ECM measurement is ~40 nm, which is ~1.6 times the elastic mean free path for this alloy,[5] so plural scattering is a concern.

Experimentally, the diameter of the nanowire has limited effect on the results over the experimentally relevant range. Figure S4 shows the measured relaxation time in the center of nanowires of thickness ranging from 28 to 60 nm. Within the uncertainty of measurement, there is no significant change in the relaxation time for thicknesses between 28 and 45 nm. When thickness increases to 60 nm, the relaxation time becomes shrinks by up to 30%. All of the experimental data were acquired from nanowires within the thickness range that yields constant relaxation time in Figure S4.

We can understand the thickness independent nature of the ECM results by considering the number of atoms that contribute to the speckle intensity at each position. Within the column approximation, each speckle in the image is determined by the atoms in a tube with diameter set by the resolution (0.7 nm for these experiments) and running through the sample thickness. That column contains ~1050 atoms, but they only contribute to the speckle intensity if they are arranged and oriented so as to create diffraction through the objective aperture superimposed on the diffraction pattern. We can estimate the likelihood of such as event by approximating the column as being comprised of randomly oriented, nanocrystalline clusters, following the model developed by Stratton and Voyles.[6] If the clusters are randomly oriented, only clusters which satisfy a Bragg condition at the particular **k** selected by the objective aperture contribute to the speckle intensity in the dark field image. We define $A$ as the fraction of clusters oriented such that there is strong diffraction. For a dark-field image,[6]

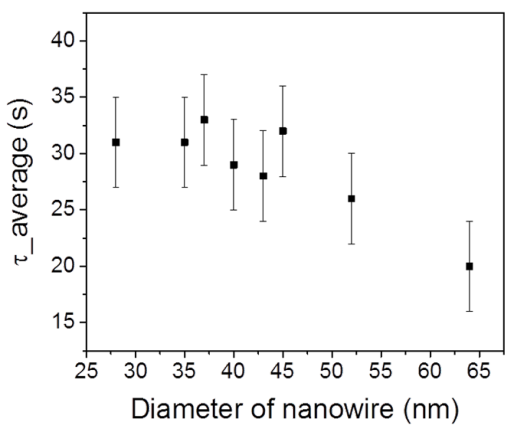

Figure S4. Average relaxation time from inner bulk of nanowire as a function of diameter of nanowire. The uncertainty is calculated by the standard deviation of the mean from five different measurements.

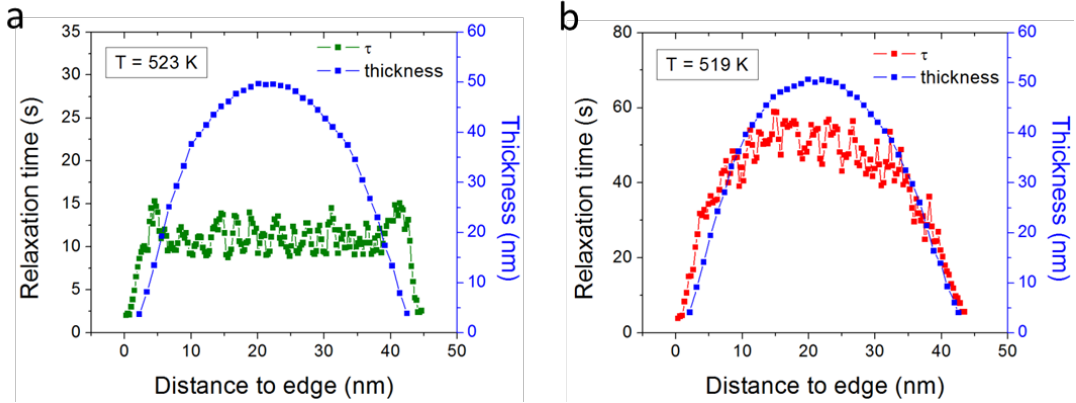

Figure S5. The thickness dependence of relaxation time from round nanowires at temperature (a) 523 K (b) 519 K.

$$A = \frac{\varphi \Delta\theta(M/2)}{4\pi} \qquad (6)$$

in which $\varphi = d_{obj}/\pi d_{diff}$, $d_{obj} = 2.51$ nm$^{-1}$ is the diameter of objective aperture in reciprocal space, and $d_{diff} = 9.43$ nm$^{-1}$ is the distance of the aperture from the direct beam of the diffraction pattern. ($\varphi$ is $2\pi$ for hollow cone dark field.) $\Delta\theta$ is acceptance angle around the exact Bragg condition which will still result in strong diffraction. For small clusters, $\Delta\theta$ is dominated by shape broadening of the reciprocal lattice spots, so $\Delta\theta = d_{atomic}/d$, in which $d_{atomic}$ is the average atomic distance 0.21 nm, and $d$ is the diameter the cluster. $M$ is the multiplicity of a family of planes in the crystal. We $d = 1$ nm as an upper bound and $M = 12$ as a typical value for low-index planes in metallic crystals.

Under these assumptions, the fraction of randomly oriented clusters contributing to the intensity of a speckle is ~5%. Essentially, while there are many clusters in real space, their scattering is well-separated in reciprocal space, so we measure only on a small fraction of them at any given time. If the entire column volume is occupied by clusters, there are ~40 clusters in each column, and around 2 of them contribute to each speckle in the image. However, in the liquid it seems likely that the density of regions with sufficient order to create diffraction is less than 100%, so we estimate that the number of clusters per speckle is 1-2 for most speckles most of the time. Similar arguments for smaller clusters and higher spatial resolution yield a similar contribution of ~1 cluster in Ångstrom beam electron diffraction experiments.[7] They also underlie the interpretation of XPCS data in terms of localized scattering from ordered regions in polymeric and colloidal liquids (*e.g.* [8]).

The shape of the nanowire does have an effect on the results. Figure S5 shows the profile of the relaxation time and projected wire thickness for a nanowire measured at 523 K and a nanowire measured at 519 K. The thickness was measured using the log-ratio method applied to elastic scattering and an elastic mean free path of 25 nm.[5] At both temperatures, the relaxation time increases in the nanowire edge, following the thickness, until it reaches a plateau in the center of the wire. These results show that the there is a thickness dependence to the data for smaller thicknesses than shown in Figure S4. Whether this is an artifact of the ECM experiments or intrinsic to the dynamics in a constrained system will be the subject of future research. At 523K, the plateau of constant relaxation time is ~30 nm wide, but at 519 K, the plateau is only 20 nm wide. This suggests that thickness effects become stronger at lower temperature, possibly as the dynamics become more spatially heterogeneous. In all cases, mean relaxation times and four point correlation functions described as "bulk" were taken within the plateau region at the center of the nanowire.

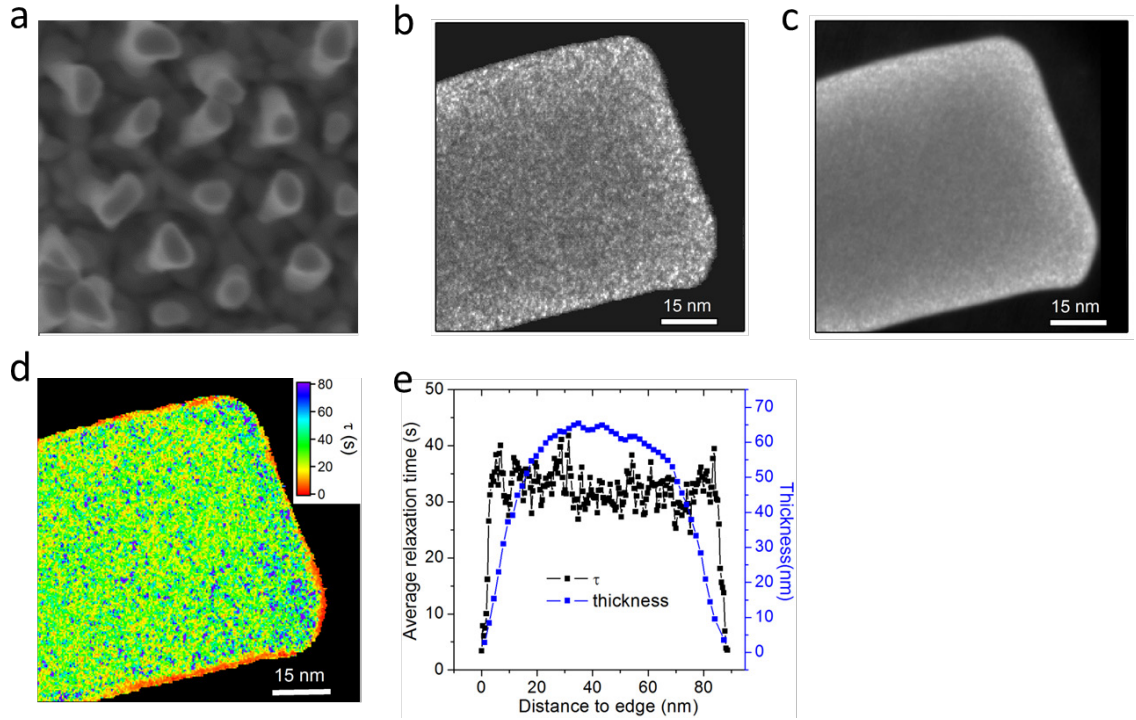

Figure S6. (a) SEM image of nanowires attached to the substrate, viewed from the top down. (b) A snapshot of dark field image movie for a rectangular nanowire at T = 519 K. (c) The time-averaged dark field image taken from the same data series in (b). (d) The spatial map of relaxation time from the same data. (e) A profile of relaxation time and thickness across the wire.

We report in the main text the existence of a near-surface region with faster dynamics than in the bulk. The data in Figure S6 show that this observation is not an artifact of the small projected thickness at the edge of the nanowires. As shown in the SEM image in Figure S6(a), some of the nanowires on the substrate have a rectangular, rather than cylindrical cross section. Figure S6(b) shows a single frame dark field image obtained from such a wire at 519 K and Figure S6(c) shows a time-averaged dark field image from the same ECM data as Figure S6(b). The residual contrast in Figure S6(b) arises from small thickness variations. Figure S6(d) is the relaxation time map derived from the entire data series. Figure S6(e) shows a thickness and relaxation time profile across part of the wire. The fast near surface layer is preserved around the edges of the wire, even when the edges are very close in thickness to the center of the wire. Thus, it is an intrinsic phenomenon, not a geometric artifact. The dynamics inside the nanowire also are consistent with measurements on round wires, showing similar spatially heterogeneous dynamics and the same mean relaxation time within experimental uncertainty.

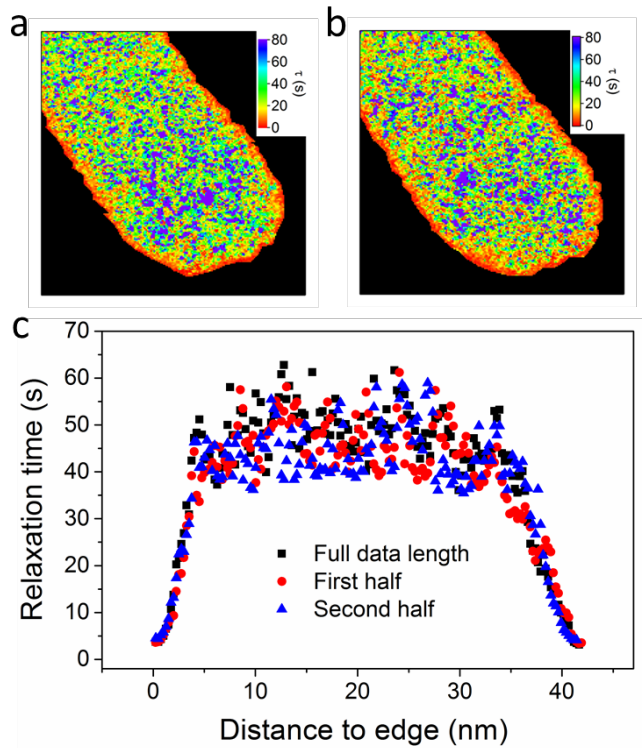

Figure S7. Spatial map of relaxation time analyzed from (a) first 1000 s (b) second 1000 s at 519 K. The time per frame is 0.25 s. (c) The average relaxation time from free surface to inner bulk from first half of the data, the second half of the data and the full length of the data.

2. Data trajectory length and effect of the electron beam

ECM experiment was performed on a nanowire at 519 K with trajectory length of 2000 s and time interval of 0.25 s, ~ 40$\tau_{med}$. Figure S7(a) and (b) show two spatial maps of relaxation time extracted from the first 20$\tau_{med}$ and second 20$\tau_{med}$ of the full data length respectively, and Figure S7(c) shows a line profile across the nanowire. The pattern of domains is not the same in the first half and second half of the data series, but the mean relaxation times are the same and the line profiles are similar. The average relaxation time from the inner bulk is 46.5 s from the first half of the series, 45 s from the second half, and 49.5 s from the correlation function calculated from the entire time series. This difference is within the ±6 s uncertainty calculated standard deviation of the mean from several similar measurements. This comparison demonstrates first that a trajectory length of 20$\tau_{med}$ is sufficient for our experiments, and second that there is no detectable change in the dynamics with exposure to the electron beam or to the microscope vacuum at elevated temperature.

3. Correlation between $\tau$ and $\beta$

Figure S8(a) shows the $\beta$ map for the same data set as the $\tau$ map in Figure 2(a) at 523 K. $\beta$ ranges from 0.2 to 1. Figure S8(b) shows $\tau$ vs $\beta$ for all the pixels in the map. There is no strong correlation between $\tau$ and $\beta$ in this data set, and the parallel graphs at other temperatures are similar.

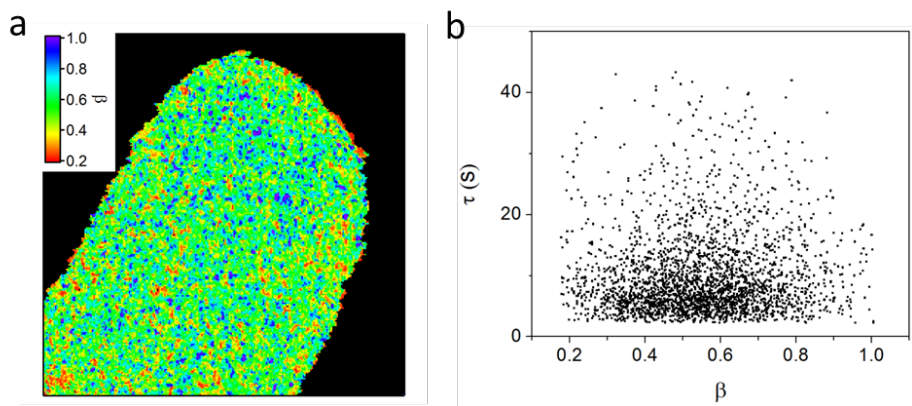

Figure S8. (a) Spatial map of stretching exponent β at 523 K. (b) τ vs β map at 523 K.

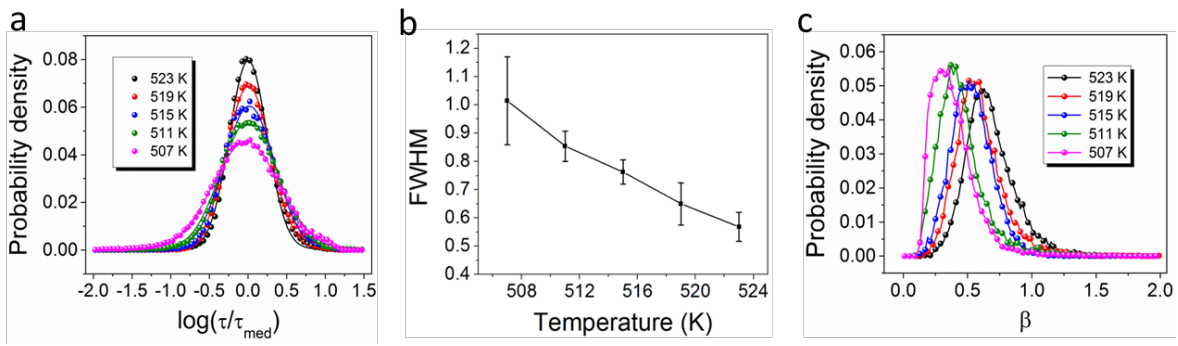

Figure S9. (a) The distributions of relaxation time normalized by the median values in log scale at all five temperatures. The solid lines are the fit. (b) The full width of half maximum acquired from the fit in (a). (c) Distributions of β shift to lower value with decreasing temperature.

4. Distribution of relaxation times and stretching exponents

Figure S9(a) shows histograms of the bulk structural relaxation time τ and β for all measured temperatures. The τ histogram $P(\tau)$ is well fit by a log-normal distribution,

$$P(\tau) = \frac{1}{\sqrt{2\pi}\sigma\tau} e^{\frac{-([\ln(\tau/\tau_{med})])^2}{2\sigma^2}}, \qquad (5)$$

where $\tau_{med}$ is the median of relaxation time and σ controls the width of the distribution. Random, non-negative processes often follow a log-normal distribution, and similar distributions have been obtained from tracer molecule rotational diffusion measurements in *ortho*-terphenyl.[9] With decreasing temperature, the full width at half maximum (FWHM) of the histograms increases, as shown as Figure S9(b). This increase in the spread of relaxation times is consistent with an increase in spatially heterogeneous dynamics as the temperature approaches $T_g$.

Figure S9(c) shows the β histograms as a function of temperature. With only a few exceptions, β is between 0 and 1. The exceptions are in some cases poor fits to the KWW form, either due to noise or due to extremely slow dynamics resulting in poorly convergence $g_2(t)$. With decreasing temperature, the β distribution shifts to smaller and smaller values. Small β is also associated with heterogeneous dynamics.

5. Temporal evolution of spatially heterogeneous dynamics

The spatial domains in the liquid state dynamics should themselves evolve with time. Wide field single molecule imaging measurements on tracer molecules in supercooled glycerol determined a persistence time for domains with relaxation time $\tau$ of ~$50\tau$ [11]. We recorded a very long data series with total data length of ~$80\tau_{med}$ was recorded with time per frame of 0.1 s at 523 K. We calculated the spatial map of $\tau$ from subsets of the time series ~$20\tau_{med}$ long, as a function of the starting time of the subset. Video 2 is the resulting "sliding window" movie of $\tau$ as a function of time and space. More quantitative analysis will be the subject of future work, but the movie shows that the domains with different dynamics fluctuate over time both in position on the wire and in relaxation time.

**Supplemental Video Captions**

Video 1(a) Time series of dark-field images acquired at 523 K. Playback is at a rate of 16 frames per second, but the data were acquired at 10 frames per second.

Video 1(b) The time series in Video 1(a) after alignment. The intensity of each speckle fluctuates, but the outline of the wire is stationary.

Video 2. Time evolution of domains with different relaxation times at 523 K, calculated using a sliding time window through a long data series. The data length for the window is ~$20\tau_{med}$ and the window interval is 20 s. The acquisition time for each frame is 0.5 s.